# Characterization of Electron Density of States in Laser-superposed Channeling Regime

**Vesna Berec**[1a, b]

[a] *Department of Physics, University of Belgrade, St. Trg 14-16, 1100, Serbia.*

[b] *Institute of Nuclear Sciences, Vinča, P. O. Box 522, 11001 Belgrade, Serbia.*

---

[1] e-mail: bervesn@gmail.com

# Characterization of Electron Density of States in Laser-superposed Channeling Regime

**Abstract**

We present low-dimensional functionalization and characterization of electron density of states (DOS) using highly correlated/precisely guided proton beam trajectories and a silicon nanocrystal as a target, representing at a same time a versatile nanolaser technique capable for coherent control of atomic quantum states and for scanning the interior of an atom with resolution comparable to 10% of the Bohr radius.

## 1. INTRODUCTION

Recently, we have described the quantum model for the subatomic confinement, Berec (2012), of the proton beam, channeled along the axial <100> axis of a Si nanocrystal (n-c). By measuring the transverse projection of electron densities inside nanocrystal we have analyzed the cross-section for the process of proton induced X-ray emission with MeV proton energies as a function of the proton impact parameters. Obtained values for the proton beam peak half-width have shown smaller diameter in comparison to size of the Bohr radius, i.e., the average radius of a hydrogen atom in its ground state. The effect of super-focusing of ions induced by channeling through thin crystals with less than 100 nm thicknesses was generated using the model of cylindrically-symmetric and harmonic continuum proton-crystal interaction potential, Gemmell (1974).

In this article, for the first time we demonstrate that the full three-dimensional structure of a single atomic orbital, Zuo *et al.* (1999), can be imaged by using the high harmonics generated from intense picosecond proton pulses focused relative to low index axial axis of a silicon nanocrystal. The model is based on the "hyperchanneling effect" of protons, Gemmell (1974), corresponding to study of Appleton *et al.* (1972) who presented the angular phase space projections in the axial hyperchanneling regime, and it is in conjunction to recent experimental investigation obtained for ultra-thin silicon crystal, Motapothula *et al.* (2012a), (2012b) .

The hyperchanneling effect characterizes the harmonic and highly correlated trajectories of swift ions (protons), which incident upon a crystal parallel to its small index symmetry axes exhibiting a series of soft small-angle scatterings that preserve the motion between the lattice rows. Namely, when charged particles enter the nanocrystal channel they produce a transverse kinetic energy determined by the potential at their point of entry and the incident angle, $\varphi$. Consequently, the region which an particle (ion, proton) can sample is determined by the condition that the net continuum potential at any point $r$ in the transverse plane, U($r$), has to be smaller than the ion transverse energy, $E_\perp$, i.e., the particle trajectory is restricted to that portion of the transverse plane, $A(E_\perp)$ where U($r$) ≤ $E_\perp$.

Conversely, the ion channeling transmission is strictly defined for the state: $U(r) > E_\perp$. The remainder of presented study deals with specific solutions of the area of the available space: $A(E_\perp) = A(r, E, \varphi)$, attainable for different conditions of well channeled transmitted protons where the proton beam is perfectly collimated and relative to Si n-c low index axis, occupying the radius of the channel center that is comparable to size of the Bohr radius.

Here we present principal study based on numerical modeling and real atomistic simulation model performed by the well known and used FLUX code (Smulders & Boerma 1987) to describe beam transmission of proton electromagnetic field in the nanocrystalline medium incorporating the feedback of electron densities under intense laser pulses. Numerically, Monte Carlo code FLUX during the initialization stage stage assigns precursor position of coordinates over the transverse plane as random distributed for an incoming beam, so that each particle can be initialized with equal probability in its accessible region; as a result the flux redistribution is calculated as a function of the angle of incidence for every point in the channel. The sizes of a bin along the *x* and *y* axes of the channel are equal to 0.005 nm. The ions backscattered from a foreign atom located (inserted) in the center of a channel will have a maximum of flux enhancement for perfect alignment ($\varphi = 0$) that will decrease with increasing $\varphi$ as the flux distributions approaches the random value.

In contrary to "standard" flux peaking effect (FPE), Kumakhov (1972a), Kumakhov (1972b), which refers to spatial focusing of channeled particles in the center of the channel (axial/planar) at λ/4 oscillation wavelength where the flux enhancement reaches only few times compared to a random level in the unit cell, the qualitatively different effect, which is specific for axial channeling, called the super-focusing effect (Demkov & Meyer 2004) predicts that the flux density can reach a thousand times higher value than the initial flux outside of the lattice. In particular, FWHM of the super-focused peak is much smaller than the thermal vibration amplitude ($\rho_{th}$ = 0.0078 nm) of the Si lattice atoms.

However, the phenomena of axial peak-focusing was not fully explained until Demkov and Meyer

(2004) performed quantum mechanical model calculations in the transverse phase space of axial channels and revisited the flux peaking effect studies. In their theoretical study it was predicted appearance of an extremely narrow, less than 0.005 nm focusing region of a highly collimated 1 MeV proton beam inside the <001> silicon channel. Study was based on the continuum proton-silicon interaction potential, dominantly cylindrically symmetric and harmonic. Assuming the calculated flux density value is sensitive to the accurate potential model, they applied Hartree-Fock approximation. The estimated FWHM of the proton beam at the super-focusing point is about 11 times smaller than the Bohr radius (0.0529 nm), i.e. the super-focusing region is less than 0.0048 nm, while the total length of the focusing region is about 17.2 nm. In another words, it was found out that beyond the possibility to locate the atom (impurity) like in standard FPE, in the case of axial flux super-focusing one can probe interior subatomic structure of a target atom.

Taking into account that the super-focused regions are pm-sized spots, until now, experimental capture of perfectly collimated beams, originating from the super-focusing, at the exit of the crystal was difficult. Very recently, Motapothula *et al*. (2012*a,b*), Motapothula (2013), demonstrated successful experimental observation of those super-focused channeled proton angular distributions in transmission regime and found excellent agreement with corresponding simulations, providing the proof for super-focusing effect existence. They captured the exit angular distributions in range from 500 keV to 2 MeV energy for protons via the 55nm [001] Si membrane using the special technique of high-resolution ion channeling, new fabrication procedure, Dang *et al.* (2013), to create ultra-thin crystal that allows uniform thickness over the entire surface, and nuclear microprobe to perform experiments on such membranes.

In our high-resolution proton scattering, we have established a hyperchanneling regime at 4 K for the proton beam and a Si nanocrystal as the target, allowing the super-focused protons to generate a nano-laser beam scans of the interior of a foreign atom that is a part of the silicon nanolattice, Pavesi & Lockwood (2004); Liang & Bowers (2010). We have used the combination of low excitation intensity, 3 and 3.5 MeV initial proton energy and the proton beam incident angles in range from zero up to 20 % of the critical angle for channeling, Lindhard (1965).

## 3. THEORETICAL AND EXPERIMENTAL MODEL

In this section, a general description of the experimental procedure and the simulation parameters, are given. To describe inner-shell ionization and excitation of atoms as a function of the impact parameter, Gemmell (1974), Appleton *et al.* (1972), we have applied semi-classical method of the coupled-channel calculations, Bransden *et al.* (1985), Thompson (1988), where the projectile following a classical trajectory provides a time-dependent electrostatic perturbation on the target electrons. A set of first-order ordinary coupled differential equations for the coefficients originating from the expansion of Molière approximation of the Thomas-Fermi (TF) interaction potential, Lindhard (1965), Gemmell (1974), the so-called coupled-channel equations, were numerically integrated along the classical trajectory of the projectile (proton) for a given impact parameters $\varphi_x$ and $\varphi_y$ that are the $x-y$ components of the proton scattering angle. The channeled proton distributions are double-mapped in the phase space: first in the coordination plane, $x'-y'$, then in the scattering-angle plane, $\vartheta_x - \theta_y$, Berec (2012), in accordance with chosen value of the crystal thickness, L and the tilt angle, $\varphi$. Nanocrystal potential between two $i, j$ lattice sites is

$$\Phi_{i,j} = U_i(r) + \frac{B}{r_{ij}^n}, \qquad (1)$$

where *n* denotes Born exponent, and coefficients *B* and *n* are experimentally obtained from the ion compressibility fitting parameters, Kittel (1995). Degree of the confinement field, *C*, and discretized energy, $\varepsilon$, respectively, are

$$C = (1+\omega_e/\Omega)^{\frac{1}{2}} s, \quad \varepsilon = (s/a) f_r \hbar \omega_e, \tag{2}$$

where $\omega_e = (4\pi e^2 n_e / m_e)^{1/2}$ is the angular frequency of the oscillations of the electron gas induced by the proton, Krause *et al.* (1994), Krause *et al.* (1996), $a = a_0 \left[ 9\pi^2 / (128 Z_2) \right]^{1/3}$ is the screening radius of the atom, with $s$ being the distance from local stability point in the harmonic continuum potential, Lindhard (1965), Gemmell (1974), and $a_0$ = 52.9 pm is the Bohr radius, Krause *et al.* (1994), $f_r$ is the average frequency of transverse proton motion close to the channel axis and $\hbar$ is the reduced Planck constant. Dispersions of the components of the proton scattering angle are $\Omega_x^2 = \Omega_y^2 = \Omega^2/2$. Details of the atomic orbital coupled-channel calculations (AO) may be found elsewhere, Thompson (1988).

When a high-intensity laser pulse interacts with collective electron plasma excitations in solid target, a several tens of percent of the plasma electrons in the vicinity of the laser focus can be effectively accelerated, forming a *directed* stream of electrons due to electrostatic fields which arise from the ponderomotive expulsion of plasma electrons from regions of high laser intensities. The acceleration process occurs within the laser pulse duration only. To obtain much stronger acceleration at the sharp density interface on the back of the nanocrystal target than on the front, we consider the proton beam energy transferred under T = 4 K to the silicon nanocrystal target (150 nm in diameter multiplied by 100 nm thickness). The induced transfer of the hot electron mean energy E = 0.7 MeV/nm is superimposed to channeled proton (CP) field, which is relative to specific nanocrystal tilts (in range: 0-20% of the critical angle for channeling), providing the transverse projection of electron DOS that allows tomographic reconstruction of the highest occupied atomic orbital of host Si. Such nano-scale precision scanning technique is based on dynamical analysis of an electron wave packet by creating a highly resolved image in real conditions with produced detailed map of discrete inter-atom positions, Berec (2012).

Recently, Dang *et al.* (2011) and Motapothula, *et al.* (2012*a,b*) performed standard axial ion channeling experiment in perfectly, crystalline, nm-thick silicon membrane close to conditions treated here, confirming the strong reduction of multiple scattering effects for ultra-thin targets in range up to 100 nm. They showed even better resolution in experimentally obtained channeled patterns in comparison to the same obtained with FLUX simulation using Ziegler-Biersack-Littmark (ZBL) Universal interaction potential.

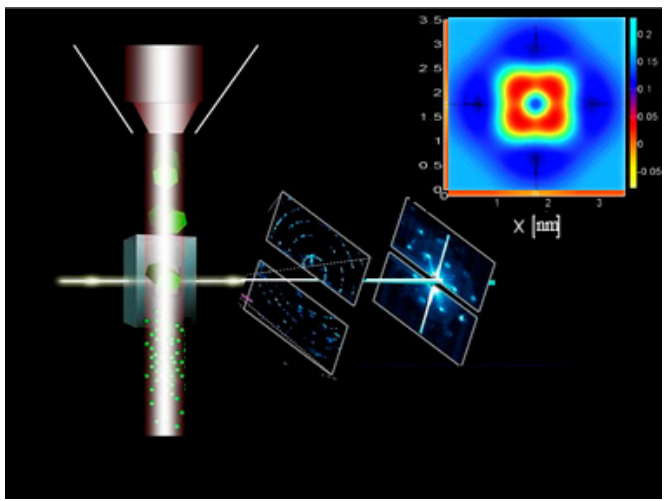

**Figure 1.** (Color online) A schematic representation of the measurement setup: laser system (vertical) and perfectly collimated (perpendicular) proton beam.
Right – top corner: tomographic result of electron DOS obtained for the considered silicon n-c target.

Similarly to former study we have performed a numerical probe using quantum Monte Carlo FLUX code, but unlike to Dang *et al.* (2011) and Motapothula *et al.* (2012*a,b*) we used the Thomas - Fermi - Molière (TFM) interaction potential corresponding to "proper" axial channeling, Lindhard (1965), i.e., the hyperchanneling – attainable for zero degree proton beam alignment toward <111> Si axis, including the tilt angles of the proton beam: $\varphi = 0.05\psi_c$, $\varphi = 0.10\psi_c$, $\varphi = 0.15\psi_c$ and $\varphi = 0.20\psi_c$ for n-c thickness of L = 100 nm.

In order to illustrate dominance and spatial confinement effect of the above potentials at various channel points we have compared the influences of the TMF - interaction potential and the ZBL Universal potential inside <111> Si channel area considering different radial atomic distances and gradual change of the tilt angle (with step 0.05°) as given in figure 2. Marked points designate the <111> channel area where the corresponding potentials exhibit the strongest confinement effect following the tilt.

In the former analysis, as well to demonstrate how the bound-state wave-packet motion appears in the harmonic spectra we use the TF proton-atom interaction potential in the Moliére approximation, Gemmell (1974), in the following form:

$$V(r') = \frac{Z_1 Z_2 e^2}{r'} \begin{bmatrix} 0.35\exp(-br') \\ +0.55\exp(-4br') \\ +0.10\exp(-20br') \end{bmatrix}, \qquad (3)$$

where $Z_1$ and $Z_2$ are the atomic numbers of the hydrogen and silicon atoms, respectively, $e$ is the elementary charge, $r'$ is the distance between the proton and atom, $b = 0.3/a$. By varying the initial tilts of the proton beam, we performed a numerical pump–probe measurement of the electron wave-packet motion, Berec (2012). We further applied the continuum approximation, Lindhard (1965), Gemmell (1974),

$$U_i^{th}(x,y) = U_i(x,y) + \frac{\sigma_{th}^2}{2}\left[\partial_{xx}U_i(x,y) + \partial_{yy}U_i(x,y)\right], \qquad (4)$$

where $U_i(x,y)$ is the continuum interaction potential of the proton and ith atomic string, which includes the effect of thermal vibrations. $\sigma_{th}$ is the one-dimensional thermal vibration amplitude of the atoms.

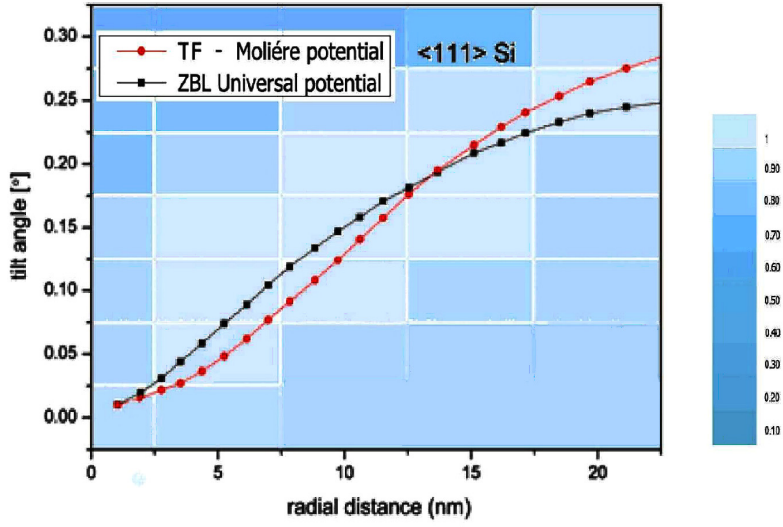

**Figure 2.** (Color online) Influences of the TFM - interaction potential (red circles) and ZBL Universal - interaction potential (black circles) on the perfectly collimated proton beam, taken for <111> axial channel of 100 nm thick Si, according to different radial distances from the atomic strings (lying on three nearest square coordination lines of the <111> axial channel) and various tilt angles. Bright areas (intensities are shown in normalized color scale) show the channel area where the corresponding potentials exhibit the strongest confinement effect.

## 2.1. Energy loss

The energy loss of the proton in the channel is taken into account via relation

$$-\frac{dE}{dz} = \frac{4\pi Z_1^2 e^4}{m_e v^2} n_e \ln\frac{2 m_e v^2}{\hbar \omega_e}, \tag{5}$$

where $m_e$ is the electron mass, $v$ the proton velocity, $n_e = \Delta U_{th}/4\pi$ is the density of the crystal's electron gas averaged along the $z$ axis, $\Delta = \partial_{xx} + \partial_{yy}$, $\Delta U_{th}$ is continuum potential of the crystal. In the MeV regime, the main contribution to the energy loss comes from inelastic interactions with electrons (Smolders & Boerma 1987). We have followed the study of Appleton *et al.* (1967) who confirmed experimentally and numerically that *K*-shell excitation in Si (in case of well channeled protons 3-11 MeV) can be neglected, while the energy loss due to ionization of the *L* shell electrons is greatly reduced (90%) for the best channeled protons in the (111) direction in Si, keeping in mind that maximal impact parameter, $b_{max}$, is ~1.6 Å. Former results imply that energy loss in particular can be fully attributed to the interaction of channeled protons with the valence electrons treated as a Fermi gas, Gemmell (1974). Hence, if one assumes the approach from Pines (1964), the energy loss due to valence electrons can be separated on two contributions coming from collective (plasma) excitations and the single particle excitations, that is:

$$-\left(\frac{dE}{dz}\right)_{val} = \frac{4\pi Z_1^2 e^4}{m_e v^2} N \left( Z_{val} \ln \frac{v}{v_F} + Z_{loc} \ln \frac{2 m_e v v_F}{\hbar \omega_p} \right), \qquad (6)$$

where $v_F$ is the Fermi velocity of the valence electron gas, $v$ is the channeled particle velocity, $NZ_{val}$ and $NZ_{loc}$ are effective electron densities for plasma and single-particle excitations, respectively, $\omega_p$ denotes the electron plasma frequency defined in presented Eq. (5) as $\omega_e = (4\pi e^2 n_e / m_e)^{1/2}$. Experimental measurement in Si, Appleton *et al.* (1967), which we followed, confirmed that the local electron density sampled by the hyperchanneled proton approximately equals to the valence electron gas density, i.e., $Z_{loc} \approx Z_{val}$. Former equation, after the substitution of $Z_{loc} \approx Z_{val} = Z$, reduces exactly to Eq. 5 where $n_e = NZ$ is density of electron gas, coinciding with the Lindhard (1965) and Winter expression, Gemmell (1974). The specific change of the dispersion of the proton scattering angle Gemmell (1974) caused by its collisions with the electrons is included in the form

$$\frac{d\Omega^2}{dz} = \frac{m_e}{m_p^2 v^2} \left( -\frac{dE}{dz} \right), \qquad (7)$$

where $m_p$ denotes the proton mass.

Note that we have taken into account experimentally confirmed fact that the energy loss of MeV channeled particle (especially at large impact parameters) exhibits quit weak contribution of collisions of the incident ion with nuclei of the lattice, and as a result enfeeble effect to the temperature dependence of the sample, Gemmell (1974) and reference therein, that means that the effect of energy loss in the MeV hyperchanneling regime cannot destroy the sample. This is in agreement with series of recent experimental studies, Dang *et al.* (2011), Motapothula *et al.* (2012), which could not be performed otherwise. Former peculiarity strongly holds even in the case of dechanneling. In turn, such property is also in agreement with the results that the maximum range does not change with the temperature, Fujimoto *et al.* (1969). It is important to note that the energy loss represents a function of distance of penetration. It changes per unit length, reaching the maximum close to end of range. According to Tavernier (2010) and references therein, range for 2 MeV protons in silicon is ~ 50 μm, which is much more than the lenght of Si sample considered in our study.

Obtained hyperchanneled Gaussian distributions were rescaled around the singularities (for reduced thickness of $\Lambda = 0.25$) providing the reduction of the noise accumulation with order: $\mathcal{O}(N-1/2)$, where $N$ is the count statistic of the proper channeled, Lindhard (1965), proton trajectories.

The analysis of the proton scattering process in question is carried out via the mapping of the impact parameter plane to the transverse position plane, Berec (2012), Gemmell (1974), in accordance with the chosen values of $\varphi$ and the crystal thickness L. The corresponding reduced crystal thickness is $\Lambda = f_r \mathrm{L} / v_0$ where $f_r = 5.94 \times 10^{13}$ Hz and $v_0$ is the initial proton velocity. The nonlinear map $J_r(x, y, \varphi, \Lambda)$ transforms the original phase space point $(x, y)$ into the phase space point where the projection is measured. The solution is expressed as

$$J_r(x, y, \varphi, \Lambda) = \frac{\partial x'}{\partial x}\frac{\partial y'}{\partial y} - \frac{\partial x'}{\partial y}\frac{\partial y'}{\partial x}, \tag{8}$$

where $x$ and $y$ are the components of the proton impact parameter, i.e., the components of its initial position in the transverse position plane. Thus, equation $J_r(x,y,\varphi,\Lambda) = 0$ determines the lines in the impact parameter plane along which the proton yield is singular.

Quantum Monte-Carlo simulations of the elastic scattering trajectories are combined with coupled channel calculations to describe inner-shell ionization and excitation as a function of impact parameter. The initial proton energy $E_0$ is 3 MeV and the crystal thickness, L, is changed from 66.1 nm to 100 nm. The temperature was set to 4 K in order to reduce the thermal vibration of the lattice. The proton beam incident angle is varied in the range between 0 and $0.20\psi_c$, where $\psi_c = \left[2Z_1Z_2e^2/(dE_0)\right]^{1/2}$ = 6.09 mrad is the critical angle for channeling, with $d$ being the distance between the crystal's atoms within the atomic strings, Kittel (1995), Bransden *et al.* (1985). The one-dimensional thermal vibration amplitude of the atoms is $\sigma_{th}$ = 7.4 pm (Batterman & Chipman 1962), Slater (1964). Binary collisions are included for each iteration using the momentum approximation, and channeled particle trajectory dependence on every atom in the chosen string is in-calculated besides the continuum potential contributions of the surrounding strings. The number of crystal's atomic strings defining the <111> channel is 36, i.e. the atomic strings lying on the three nearest square coordination lines, Kittel (1995), Krause *et al.* (1994), are considered. The implicit Runge–Kutta method of the fourth order, (Abramowitz & Stegun 1965), (Parker & Chua 1989), is used to iterate the solutions of protons trajectories from the uniform distributions within the region of the channel. The components of the proton impact parameter, $x$ and $y$, are mapped into single density matrix where the initial number of protons was set to $4 \times 10^7$.

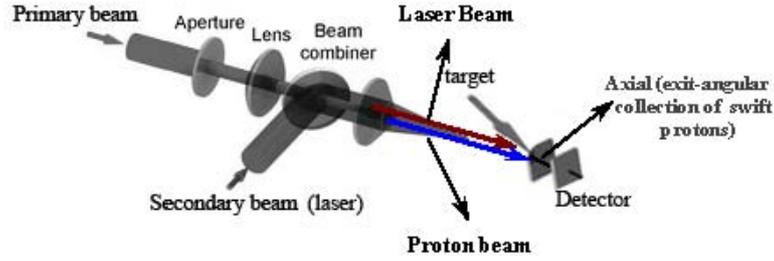

**Figure 3.** (color online) A schematic of the ''standard'' experimental setup, representing the main part of the beam-line focused with a spherical lens to the target position and exit plane axial collection of the swift protons, with a pierced mirror and a spherical lens on entrance of a fiber optic coupled with the detector. After passing the beam combiner, the laser beam (red) and the proton beam (blue) are spatially separated by the critical angle for <111> Si: $\varphi = \psi_c$.

Laser system (represented schematically in figures 1 and 3) induces optimization of the potential barrier (in target) increasing it compared to the transverse energy $E_\perp$ of the channeled proton. As a result, the proton motion is additionally stabilized and supplementary confined in interatomic potential area located outside to valence electron distances Barrett *et al.* (1975), i.e., where the energy loss is anomalously low. Besides, by setting up the intense field at the rate close to $a_0 = 1$ au which binds over the Coulomb force - creating in such way the laser-dressed atoms, which are proved to be stable toward energy losses induced by ionization, Gavrila (2002), it is possible to induce the atomic stabilization in target. Laser stabilization of atoms is a well known and experimentally confirmed, Reiss (2001), specific steady-state created under asymptotically high-frequency and high-intensity radiation field of constant amplitude, which prevents ionization of an atomic electron, Gavrila (2002). The intensities required to reach the stabilization regime are within standard tabletop amplified sub-picosecond laser systems, which can exceed $10^{18}$ W/cm$^2$ for λ = 800 nm, as in case of conventional Ti: sapphire laser.

To model numerically a nondecaying bound states in the intense laser field we have used full three-dimensional KH stabilization regime van Druten *et al.* (1997), where the KH potential are further expanded into a Fourier series

$$V(r + \alpha(t)) = \sum_n V_n(r, \theta; \alpha)\, e^{in(\varphi - \omega t)}, \tag{9}$$

where $\alpha = F_0/\omega^2$ is the pulse amplitude, and $\theta$, $\varphi$ are the polar angles defining circular Coulomb potential in short-range approximation, Tikhonova *et al.* (2002). The maximum laser induced electric-field amplitude, $E_0$, corresponds for a Ti: sapphire laser ($\lambda$ = 800 nm) to a peak intensity of $\approx 2.16 \cdot 10^{18}$ Wcm$^{-2}$, where an angular frequency of $\hbar\omega = 27.21\, eV$ is used as a standard value employed in studies of atomic stabilization.

The details of the channeling experimental setup, detection system and a sample preparation are the same as those reported in Motapothula *et al.* (2012*a,b*), however the function and versatility of experimental setup are extended by a laser system. In figure 3, a secondary collimated, laser-beam is deflected through the beam combiner and redirected toward target under small impact parameter (which corresponds to critical angle, $\psi_c$) close to atomic strings (in order to establish control over valence electron dynamics).

## 3. RESULTS

Figure 4. shows the tomographic reconstruction (in velocity phase space) of electron density of states (DOS) for the highest occupied atomic orbital in the $\pm 5$ nm focal region inside the silicon n-c core. The proton beam energies are 3 and 3.5 MeV, the incident angles are $0.05\psi_c$ and $0.15\psi_c$, and the n-c thickness, L = 100 nm. $\varphi$ = $0.05\psi_c$, $0.15\psi_c$, are the proton beam incident angles along the x (y) axis in the transverse position plane. At the 3 and 3.5 MeV excitation intensity and the proton

beam tilt angles (relative to <111> axis): $0.05\psi_c$, the spectrum contains one maximum of a single broad transmission peak FWHM of 0.482 nm.

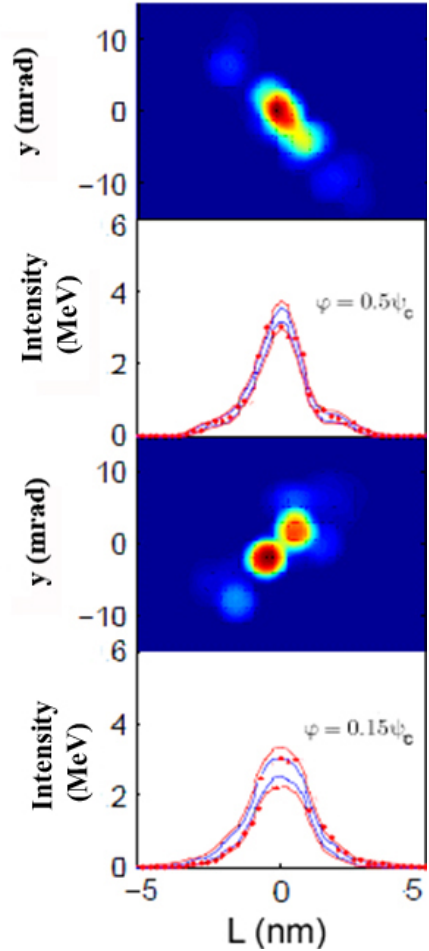

**Figure 4.** (Color online) Proton beam intensity distribution in transverse position plane and the tomographic reconstruction of the highest occupied atomic orbital representing electron density of states (DOS) in the 5 nm focal region inside the silicon n-c core, obtained by the numerical probe. Here applied TF-Molière interaction potential includes real data parameters: Debye thermal vibration amplitude ($\sigma_{th}$ = 7.4 pm), exact Young's modulus which is largest along the <111> Si axis (E = 187.5 GPa) and high mechanical quality factor of silicon ($Q > 10^7$). DOS along the $x - y$ axis are obtained in the configuration phase space and velocity phase space for L = 100 nm thick Si n-c. Two-dimensional representation of channeled proton flux is shown for the proton beam incident angles: $\varphi$ = $0.05\psi_c$, $0.15\psi_c$ (the proton beam energies are 3 and 3.5 MeV) in the transverse position plane. The corresponding Gaussian fits are red colored.

The proton transmission through the Si n-c channel close to low index axis is accompanied with spontaneous emission of approximately 140 meV below the silicon band gap that is generally assigned to the recombination of the excitons (Liang & Bowers 2005). The maximal enhancement in proton flux density corresponds to area inside the central equipotential curve, of axial crystal channel

$$\gamma\left(A_{l_0}(E_\perp)\right) \approx \ln\left|\frac{A_0 k}{\pi E \varphi^2}\right| \qquad (10)$$

where $\gamma\left(A_{l_0}(E_\perp)\right)$ denotes effective potential area as a function of the incident proton beam transverse energy, Berec (2012), Lindhard (1965).

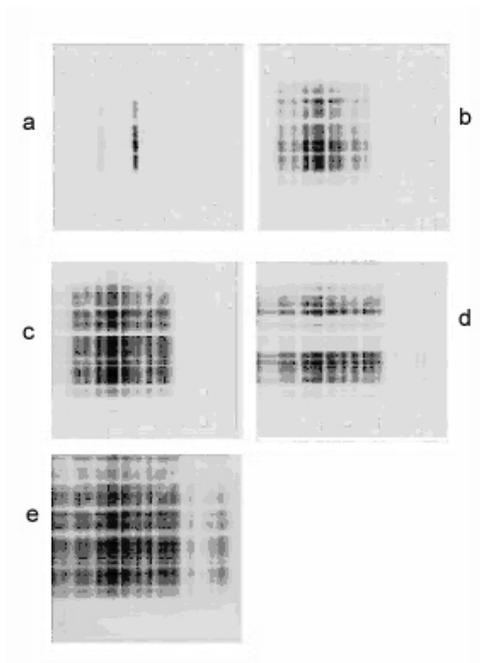

**Figure 5.** Numerically obtained transmission patterns in the exit - scattering-angle plane of 100 nm thick Si n-c. The ion beam divergence is set to 0.1 mrad. Two-dimensional representation of the hyperchanneled proton trajectories evolution with incident angles: (a) $\varphi = 0$, (b) $\varphi = 0.05\psi_c$, (c) $\varphi = 0.10\psi_c$, (d) $\varphi = 0.15\psi_c$ and (e) $\varphi = 0.20\psi_c$, corresponding to n-c thickness of L = 100 nm, are mapped from the transverse position plane (acording to Eq. 8).

The results obtained for 3 and 3.5 MeV proton beam energies and tilt angle $\varphi = 0.15\psi_c$, show the nonequilibrium density of states across the central part of the channel as a nonuniform flux redistribution. We can clearly see appearing of one central maximum as the sharply focused spot, followed by one asymmetrical pairs of lateral patterns due to strong perturbation of the harmonic potential in the channel. This implies the strong anharmonic effect on the reversibility principle for the Si n-c, in the effective continuum interaction potential, even in vicinity of low index $\langle 111 \rangle$ axis where one would expect the uniform spectrum of transverse energy.

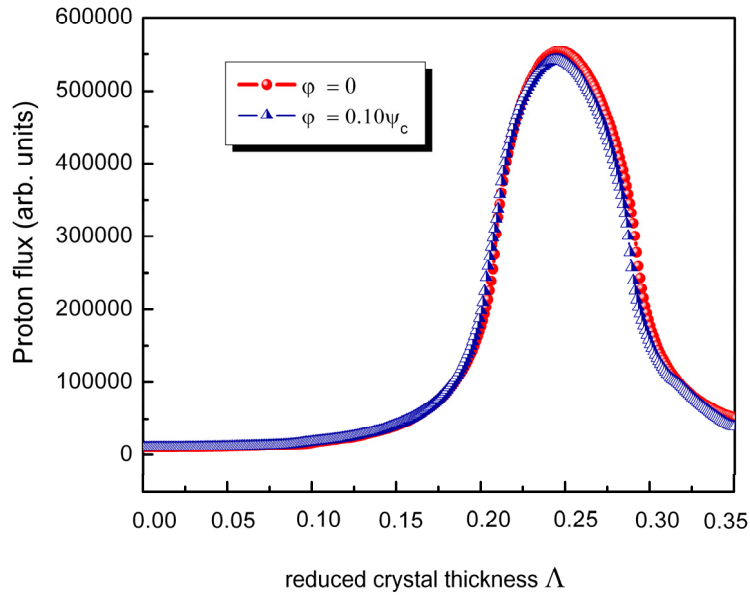

**Figure 6** (Color online) The axial yield (flux) of 2 MeV channeled protons as a function of the reduced n-c thickness in vicinity of the super-focusing point (0.25) for the tilt angles $\varphi = 0$ and $\varphi = 0.10\psi_c$ along $y$ axis in the transverse position plane, numerical simulation results obtained using the QMC FLUX code.

Figure 5 shows proton transmission patterns in the two-dimensional representation on the exit-scattering plane of the n-c, which are obtained by mapping of positions in transversal position plane, Eq. (8), for incident angles: $\varphi = 0$, $0.05\psi_c$, $0.10\psi_c$, $0.15\psi_c$ and $0.20\psi_c$, and thickness L = 100 nm. It is evident that for $\varphi = 0$, $0.05\psi_c$, $0.10\psi_c$ and $0.15\psi_c$ each of the distributions has one strong and narrow maximum, while for $\varphi = 0.20\psi_c$, proton distribution becomes broader, and fairly uniform. As the tilt angles decrease, the transmission pattern becomes narrower because of the preferential amplification of frequencies close to the maximum of the energy spectrum. It is also evident that for the values of the crystal tilt angle equal to $\varphi = 0.15$, spatial distribution structure exhibits characteristic splitting in two partitions, along horizontal axis, this is accompanied by a strong rearrangement and decrease in the proton flux. Thus, as the tilt angle decreases, the spatial distributions approach to the center of the channel, accompanied with maximum value of the channeled protons yield, for the zero-degree tilt angle. From the results in figure 4, one can determine a quantitative correlation between the maximal value of the channeled protons yield (flux) and the super-focusing effect. The increase in crystal tilts magnitude relative to 15%, leads to annihilation of the super-focusing effect, i.e., the patterns become broader, inhomogeneous, with low peak intensity. One can clearly see the resolution precision in measurement of the electron density within the foreign atom as a function of the proton impact parameters. Such scanning probe method of the interior of the foreign atom allows the subatomic microscopy.

Figure 6 gives the resolution of the proton beam focusing spot. It shows channeled protons flux evolution with the reduced crystal thickness parameter $\Lambda$ (between 0 and 0.3) in the focusing region inside the Bohr radius, around the centers of their spatial distributions, for $\varphi = 0$ and $0.10\psi_c$. A full discussion of the reduced thickness is presented elsewhere, Krause *et al.* (1994), Krause *et al.* (1986), Motapothula (2013). The full-widths of the two maxima are 0.044 and 0.043, respectively, and the corresponding values of $\Delta L$ are 27.8 and 27.5 nm, respectively. The small downward shifts from the super-focusing point at $\varphi = 0$ are attributed to small perturbation induced via anharmonic component of the continuum proton-crystal interaction potential. Taking into account lattice thermal vibration the crystal should be cooled to a temperature well below room temperature to make the effect more pronounced.

Figure 7 indicates that axial yield (flux) of the channeled protons in the transverse position plane, for the fixed value of the crystal tilt angle $\varphi = 0.10$ and different crystal thicknesses: L = 79.32 nm, L = 83 nm, L = 85.93, i.e., corresponding to points: $\Lambda = 0.24$, $\Lambda = 0.25$, $\Lambda = 0.26$, presents a strongly damped function of nanocrystal thickness with a periodic behavior. Note that, maxima and minima presented in figure 7 gradually become less prominent with increase of Si n-c thickness.

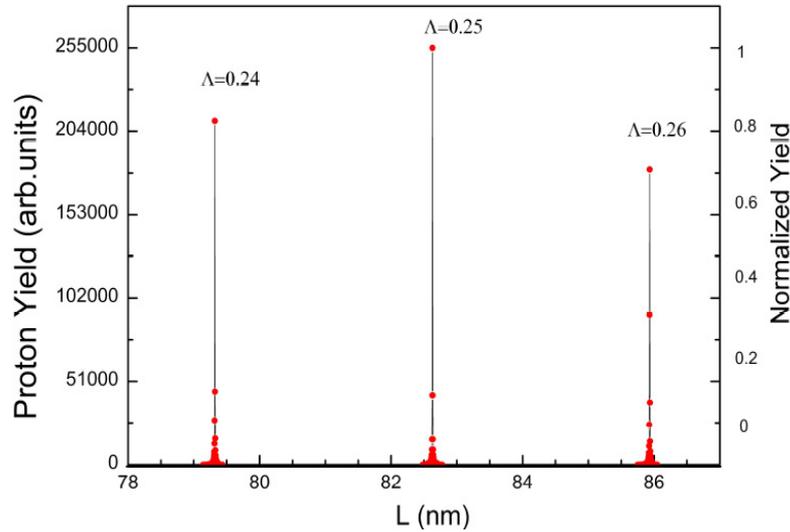

**Figure 7.** (Color online). Calculated maximal axial flux intensities for super-focused proton beam in the transverse position plane. The fixed zero tilts parameters, $\varphi = 0$, are taken in different positions relative to n-c channel <111> axis corresponding to points: $\Lambda = 0.24$, 0.25 and 0.26. The distance between the maxima of the proton yields induced by CP confinement field is calculated via $\pi^2 \hbar^2 (n+1/2)/(L^2 m_e)$, L = 100 nm, for n-discretized 2-d proton n-c crystal potential.

Further analysis inside the region: $-a_0 \leq$ L point at 83nm $\leq a_0$, shows that only FWHM of the first pronounced maxima coincide with the proton beam super-focusing point, while the second and third maxima appear as a result of passing the rainbow lines, Krause *et al.* (1994), Nye (1999), through the transverse position plane of the n-c. From the point of view of longitudinal spatial resolution, the quantitative analysis shows that proton beam obtained with n-c tilt $\varphi = 0.10$, for $\Lambda = 0.24$, $\Lambda = 0.25$ and $\Lambda = 0.26$, enters into the cylindrical region which corresponds to one tenth of the Bohr radius $0.1 a_0 = 0.0053$ nm around the channel axis, i.e. this is the analyzed length of the super-focusing area. The obtained results further confirm the focusing of protons into the region whose dimension is much smaller than the average radius of a hydrogen atom in its ground state.

Figures 8 and 9 show the FWHM and 3-D spatial projection of hyperchanneled proton beam flux distribution simulated in the transverse position plane, for zero degree n-c tilt angle.

It is important to note that only this flux peak represents singular proton yield, i.e. the flux of channeled protons has absolute maximum for $\varphi = 0$.

The subatomic scanning resolution, Breese *et al.* (1996), obtained via different proton beam intensity profiles, illustrated by figure 9, is calculated up to crystal tilt $0.10\ \psi_c$ corresponding to different longitudinal, L, and transversal, x, thicknesses.

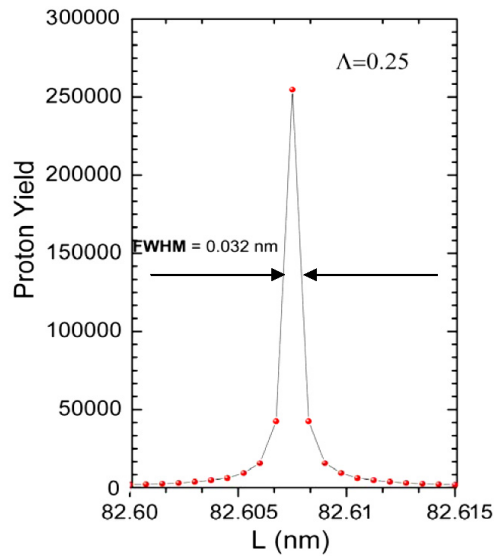

**Figure 8.** (Color online) The proton axial yield projection in the transverse position plane at the super-focusing point ($\Lambda = 0.25$), for angle $\varphi = 0$, numerical simulation results obtained using the QMC FLUX code.

Total resolution is determined by the ratio of the inserted Phosphorous atom diameter and the full-width of the proton beam. For this case, the calculated ratio is 20, which well corresponds with Slater (1964). In this case the sensitivity of the spectroscopy is determined by the way the cross-section depends on the electron waive density.

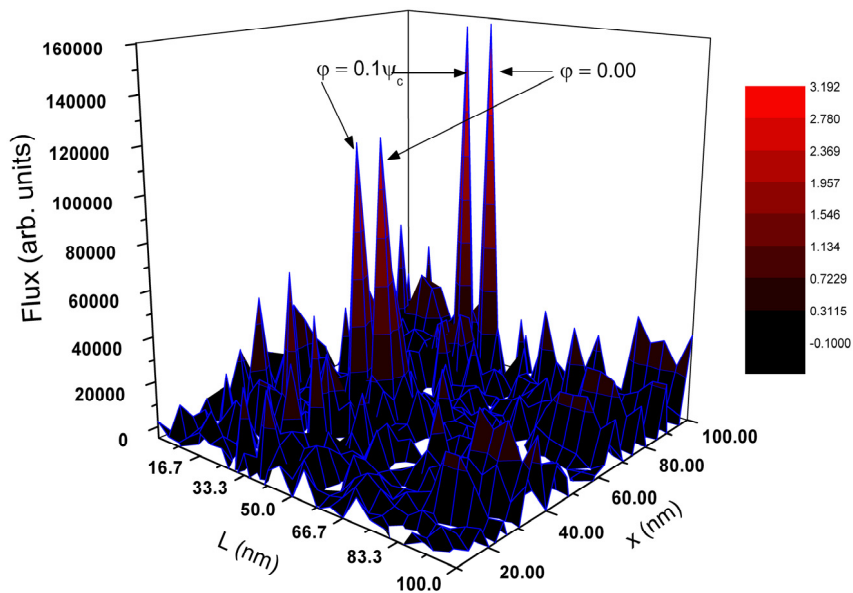

**Figure 9.** (Color online) Comparison of the proton beam intensity profiles (different transverse cross-sections) for the tilt angle $\varphi = 0.10\psi_c$ that corresponds to 10% of the critical angle for channeling, relative to super-focusing point, obtained for different longitudinal, L, and transversal, x, thickness of the nanocrystal, numerical simulation results obtained using the QMC FLUX code.

Sampling by angular, $\vartheta$, parameter (Eq. (8)) up to 20% of the critical angle is enough to resolve electron DOS, showing the nodal structures in $\langle 111 \rangle$ Si n-c channel, as can be seen in figure 10.

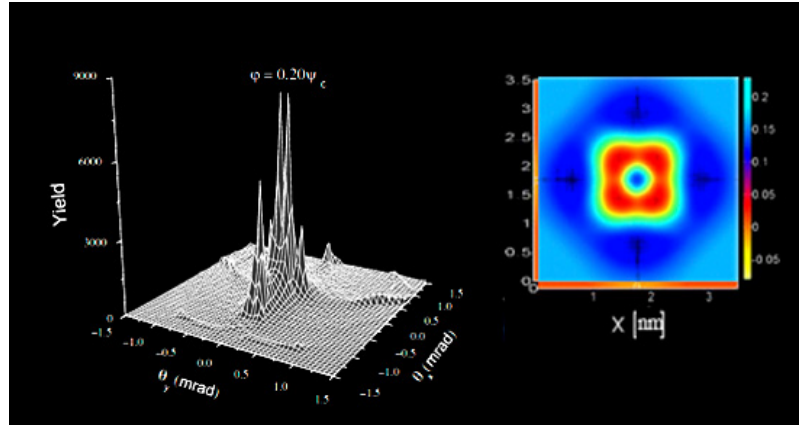

**Figure 10.** (Color online) Right: 2D Intensity profile of the highest occupied atomic orbital representing electron density of states (DOS) in $\langle 111 \rangle$ Si n-c channel obtained for the beam angle $0.20\psi_c$. Left: 3d-surface slices in angular-exit n-c plane obtained for the same angle, $0.20\psi_c$.

## 5. CONCLUSIONS

We have demonstrated, theoretically and numerically, the ability of pulse-generated hyperchanneled protons to measure the three-dimensional structure of atomic orbitals. By using the coupled-channel calculation formalism we clarified an intimate connection between the electron density of states and the anharmonic component of the proton-crystal interaction potential induced by the channeling relative to Si n-c <111> axis, and demonstrated the achievable coherent control obtained via super-focused proton beam in the Si n-c, aimed to produce a single nanolaser capable for scanning the interior of an atom within the region below the Bohr radius for all considered values of the proton beam incident angles.

**Acknowledgment**

The Author is thankful to Dr. I. Anicin on fruitful discussions. This work is supported from the Ministry of Science and Education RS.